%% file: RR-6370.tex
\documentclass[a4paper]{article}
\usepackage{RR}
\RRNo{6370} 
\usepackage{hyperref}
\usepackage{amsmath,amssymb}
\usepackage{epsfig}
\usepackage{multirow}
\usepackage{pslatex}
\usepackage{pst-node}
\usepackage{enumerate}
\usepackage[T1]{fontenc}
\usepackage[latin1]{inputenc}
\usepackage{longtable}
\usepackage{float}
\RRdate{Novembre 2007}

\RRauthor{
Rumen Andonov \thanks{IRISA and University of Rennes 1, Campus de Beaulieu, 35042 Rennes, France}
\and
Nicola Yanev \thanks{University of Sofia, Bulgaria}
\and
Noël Malod-Dognin \thanks{IRISA and University of Rennes 1, Campus de Beaulieu, 35042 Rennes, France}
}
\authorhead{R. Andonov, N. Yanev and N. Malod-Dognin}
\RRtitle{Vers une classification structurelle des protéines basée sur le recouvrement des cartes de contacts}

\RRetitle{Towards Structural Classification of Proteins based on Contact Map Overlap}
\titlehead{Towards Structural Classification of Proteins based on Contact Map Overlap}
\RRresume{
Une multitude de mesures ont été proposées pour quantifier la similarité entre les structures 3D de protéines.
Parmis ces mesures, la maximisation du recouvrement des cartes de contacts ("Contact Map Overlap Maximization", CMO) a reçu durant les dix dernières années une attention soutenue, car elle permet une bonne estimation des relations naturelles d'homologie entre protéines.
Cependant, malgré l'implication des communautés de bio-informatique et de sciences computationnelles, les performances des algorithmes connus restent modestes.
\`A cause de la complexité du problème, ces algorithmes sont limités à de petites instances et ne sont pas applicables pour des comparaions à grandes échelles.

Ce rapport marque une nette amélioration sur ce point par rapport aux méthodes précedentes.
Nous présentons un nouveau modèle de programmation linéaire en nombre entier pour CMO, et nous proposons un algorithme exact de séparation et évaluation dont les bornes proviennent de la relaxation lagrangienne de notre modèle.
L'efficacité de cette approche est démontrée sur un petit ensemble de test connu (l'ensemble de skolnick, 40 domaines). Sur ce jeu de test, notre algorithme surpasse en rapidité d'exécution les meilleurs algorithmes existants tout en obtenant des bornes de meilleurs qualité.
Quelques instances difficiles de CMO ont été résolues pour la première fois, et ce en des temps raisonnnables.
\`A partir des valeurs de temps de calculs et de "gaps" relatifs (la différence relative entre la borne supérieur et inférieure), nous avons obtenu la bonne classification de l'ensemble de skolnick.
Ces résultats encourageants nous ont poussés à créer un jeu de test plus difficile pour confirmer les capacités de classification de notre approche.
Nous avons construit un ensemble de test contenant 300 domaines de protéines (un sous-ensemble d'ASTRAL) que nous avons appelé Proteus\_300.
En utilisant le gap relatif des 44850 couples comme une mesure de similarité, nous avons obtenu une classification en très bon accord avec SCOP.
Notre algorithme offre donc un outil puissant pour la classification de grandes bases de données de structures.
}

\RRabstract{
A multitude of measures have been proposed to quantify the similarity between protein 3-D structure.
Among these measures, contact map overlap (CMO) maximization deserved sustained attention during past decade because it offers a fine estimation of the natural homology relation between proteins.
Despite this large involvement of the  bioinformatics and computer science community, the performance of known algorithms remains modest.
Due to the complexity of the problem, they got stuck on relatively small instances and are not applicable for large scale comparison.

This paper offers a clear improvement over past methods in this respect.
We present a new integer programming model for CMO and propose an exact B\&B algorithm with bounds computed by solving  Lagrangian relaxation.
The efficiency of the approach is demonstrated on a popular small benchmark (Skolnick set, 40 domains). On this set our algorithm significantly outperforms the best existing exact algorithms,  and yet provides lower and upper bounds of better quality.
Some hard CMO instances have been solved for the first time and within reasonable time limits.
From the values of the running time and the relative gap (relative difference between upper and lower bounds), we obtained the right classification for this test.
These encouraging result led us to design a harder benchmark to better assess the classification capability of our approach.
We constructed a large scale set of 300 protein domains (a subset of ASTRAL database) that we have called Proteus\_300.
Using the relative gap of any of the 44850 couples as a similarity measure, we obtained a classification in very good agreement with SCOP.
Our algorithm provides thus a powerful classification tool for large structure databases.
}

\RRmotcle{
Alignement de structures de protéines, maximisation du recouvrement de cartes de contacts, optimisation combinatoire, programmation linéaire en nombre entier, séparation et évaluation, relaxation lagrangienne.
}

\RRkeyword{
Protein structure alignment, Contact Map Overlap maximization, combinatorial optimization, integer programming, branch and bound, Lagrangian relaxation.
}
\RRprojets{SYMBIOSE}
\RRtheme{\THBio} 
 \URRennes 
\setlongtables

\newtheorem{prop}{Proposition}

\newcommand{\ba}{\begin{array}}
\newcommand{\ea}{\end{array}}

\begin{document}
\makeRR   

\tableofcontents

\section{Introduction}\label{sec:intro}
A fruitful assumption of molecular biology  is that proteins sharing close three-dimensional (3D) structures are likely to share a common function and in most case derive from a same ancestor. Computing the similarity between two protein structures is therefore a crucial task and has been extensively investigated \cite{1001,xu-gang,halperin,adam}.
Interested reader can also refers \cite{101,caprara,lancia,strickland,agarwal,krasnogor2,3djudge,krasnogor}. 
Since it is not clear what quantitative measure to use for
comparing protein structures, a multitude of measures have been
proposed. Each measure aims in capturing the intuitive notion of similarity. 
We studied the \emph{contact-map-overlap} (CMO) maximization,
a scoring scheme  first proposed in \cite{goldman}. This measure has been found to
be very useful for estimating  protein similarity - it is robust,
takes partial matching into account, translation invariant and
captures the intuitive notion of similarity very well. 
The protein's primary sequence is usually thought  as
composed of residues. Under specific physiological conditions, the
linear arrangement of residues will fold and adopt a complex 3D shape, called native state (or tertiary structure). In its native state, residues that are far away along
the linear arrangement may come into proximity in 3D space. 
The proximity relation is captured by a contact map.  Formally, a map is specified by a $0-1$
 symmetric squared  matrix $C$ where $c_{ij}=1$ if the Euclidean distance of two heavy atoms (or the minimum distance
between any two atoms belonging to those residues) from the $i$-th
and the $j$-th amino acid of a protein is smaller than a given
threshold in the protein native fold. 
In the  CMO approach one tries to evaluate the similarity of two
proteins by determining the maximum overlap (also called
alignment) of contacts map.  Formally:  given two adjacency  matrices, 
find two sub-matrices that correspond to principle minors\footnote{matrix that corresponds to a principle minor is a sub-matrix of a squared matrix obtained 
by deleting   $k$ rows  and the same $k$  columns}   
having the maximum inner product if thought as vectors (i.e. maximizing 
the number of 1 on the same position). 

The counterpart of the CMO problem in the graph theory is the well known maximum common subgraph problem (MCS) \cite{gary}. The bad news for the later is its
APX-hardness\footnote{see \text{"A compendium of NP optimization problems"}, \text{http://www.nada.kth.se/$\sim$viggo/problemlist/}}
The only difference between the above defined CMO and MCS is that
the isomorphism used for the MCS is not restricted to the
non-crossing matching only. Nevertheless the CMO is also known to be NP-hard \cite{cmo-hard}.
Thus the problem of designing efficient algorithms that guarantee the CMO quality is an important one that has eluded researchers so far.
The most promising approach for solving CMO seems to be integer programming coupled with either Lagrangian relaxation \cite{1001} or B\&B reduction technique \cite{cmos_07}.

The results in this paper confirm once more the superiority of Lagrangian relaxation  to CMO 
since the algorithm we present belongs to the same class.
Our interest in CMO was provoked by its similarity with the
protein threading problem. For the later we have presented an approach based on the so called  non-crossing matching in bipartite graphs \cite{ABY04}. It yielded a highly efficient algorithm solving the PTP by using the Lagrangian duality \cite{Bal04,cost,jcomp}.

The contributions of this paper are as follows.
We propose a new integer programming formulation of the CMO problem.
For this model, we design a B\&B algorithm coupled with a new Lagrangian relaxation for bounds computing.
We compare our approach  with the best existing exact algorithms \cite{1001,cmos_07} on a widely used benchmark (the Skolnick set), and we noticed that it outperforms them significantly.
New hard Skolnick set instances have been solved.
In addition, we observed that our Lagrangian approach produces upper and lower bounds of better quality than in \cite{1001,cmos_07}. This suggested us to use the relative gap (a function of these two bounds) as a similarity measure. To the best of our knowledge we are the first ones to propose such criterion for similarity. Our results demonstrated the very good classification potential of our method.
Its capacity as classifier was further tested on the Proteus\_300 set, a large benchmark of 300 domains that we extracted from ASTRAL-40 \cite{astral}.
We are not aware of any previous attempt to use a CMO tool on such large database.
The obtained classification is in very good agreement with SCOP classification.
This clearly demonstrates that our algorithm can be used as a tool for large scale classification.

\section{The mathematical model}\label{sec:model}

We are going to present the CMO problem as a matching problem in a bipartite
graph, which in turn will be posed as a longest augmented path problem in a
structured graph.  Toward this end we need to introduce few notations as
follows. The contacts maps of two proteins P1 and P2 are given by graphs
$G_m=(V_m,E_m)$ with $V_m=\{1,2,\ldots,n_m \}$ for $m=1,2$. The vertices $V_m$
are better seen as ordered points on a line and correspond to the residues of
the proteins. The arcs $(i,j)$ correspond to the contacts. The right and left
neighbouring of node $i$ are elements of the sets $\delta_m^+(i)= \{j | j>i,
(i,j) \in E_m \}$, $\delta_m^-(i)= \{j | j<i, (j,i) \in E_m \}$. Let $i\in V_1$
be  matched with $k \in V_2$ and  $j\in V_1$ be matched with $l \in V_2$. We
will call a matching \emph{non-crossing}, if $i<j$ implies $k<l$. A feasible
alignment of two proteins $P_1$ and $P_2$ is given by a non-crossing matching
in the complete bipartite graph $B$ with a vertex set $V_1 \cup V_2$.

Let  the weight $w_{ikjl}$ of the matching couple $(i,k)(j,l)$ be set as
follows
\begin{equation}
  w_{ikjl}  = \left\{
  \begin{array}{ll}
    1 & \mbox{ if } (i,j) \in E_1   \mbox{ and  } (k,l) \in E_2 \\
    0 & \mbox{ otherwise}
  \end{array} \right.
\end{equation}
For a given non-crossing matching $M$ in $B$ we define its weight $w(M)$ as a
sum over all couples of edges in $M$. The CMO problem consists then  in
maximizing $w(M)$, where $M$ belongs to the set of all non-crossing matching
in $B$.

In \cite{ABY04,Bal04,cost,jcomp} we have already dealt with non-crossing matching
and we have proposed a network flow presentation of similar one-to-one
mappings (in fact the mapping there was many-to-one). The
adaptation of this approach to CMO is as follows:
  The edges of the
bipartite graph $B$ are mapped to the  points of $n_1\times n_2$
rectangular grid $B'=(V', E')$ according to: point - $(i,k) \in
V'$ $\longleftrightarrow$ edge - $(i,k)$ in $B$.

\textbf{Definition}.
 The \textbf{feasible path} is an arbitrary sequence $(i_1,k_1),
(i_2,k_2), \ldots, (i_t,k_t)$  of points in $B'$ such that
$i_j<i_{j+1}$ and $k_j<k_{j+1}$ for $j=1,2,\ldots,t-1$.

\begin{figure}[ht]
\centering
\epsfig{figure=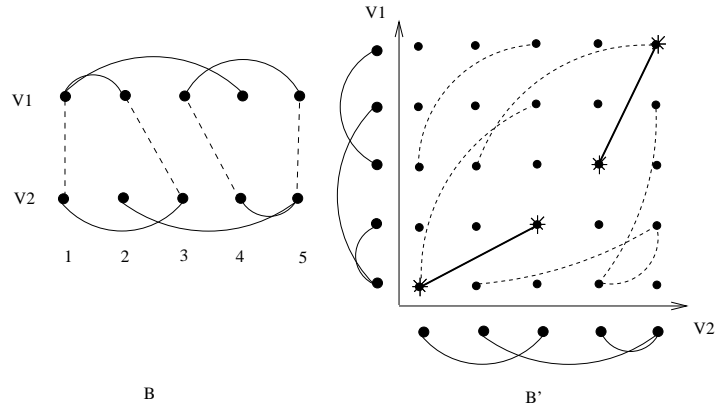,scale=.40}
\caption{Left: Vertex 1 from V1 is matched with vertex 1 from V2 and 2
is matched with 3: matching couple $(1,1)(2,3)$. Other  matching couples are
$(3,4)(5,5)$. This defines a feasible matching $M=\{(1,1)(2,3),(3,4)(5,5)\}$
with weight $w(M)=2$. Right: The same matching is visualized in graph $B'$.}
\label{fig:B}
\end{figure}

  The correspondence feasible path $\leftrightarrow$
non-crossing matching is obvious. This way non-crossing matching problems are converted to problems on feasible
paths. We also add arcs $(i,k)\rightarrow (j,l) \in E'$  iff  
$w_{ikjl}=1$. In $B'$, solving CMO corresponds to finding the
densest (in terms of arcs) subgraph of $B'$ whose node set is a
feasible path (see Fig.~\ref{fig:B}).

\begin{figure}[ht]
\centering \epsfig{figure=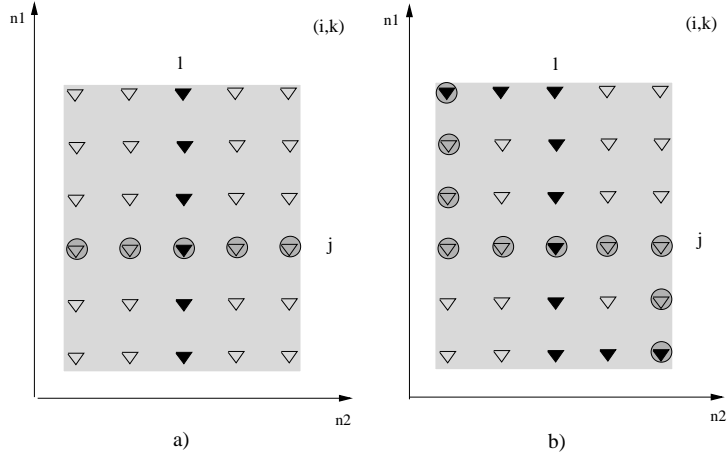, scale=.55}
\caption{ The shadowed area represents the set of vertices in $V'$ which are tails
for the arcs heading to $(i,k)$. \quad In a): $\blacktriangledown$
corresponds to the indices of $y_{jlik}$ in (\ref{const4}) for $l$
fixed.   $\bigcirc$ corresponds to the indices of $y_{jlik}$ in (\ref{const2}) for $j$ fixed.   
\quad In b): $\blacktriangledown$ corresponds to the indices
of $y_{jlik}$ in (\ref{const4a}) for $l$ fixed (the set $col_{ik}(l)$).   $\bigcirc$ corresponds to the indices of $y_{jlik}$ in (\ref{const2a}) for $j$ fixed    (the set $row_{ik}(j)$).
}\label{fig:path}
\end{figure}

To each node $(i,k) \in V'$ we associate now a $0/1$ variable
$x_{ik}$, and to each arc  $(i,k)\rightarrow (j,l) \in E'$, a
$0/1$ variable $y_{ikjl}$. Denote by $X$ the set  of feasible
paths. The problem can now be stated as  follows  (see Fig. \ref{fig:path} a)   for illustration)

\begin{equation}\label{obj}
v(CMO)=\max\sum_{(ik)(jl)\in E'} y_{ikjl}
\end{equation}
subject to
\begin{equation}\label{const1}
x_{ik}\ge\sum_{l \in \delta_2^+(k)} y_{ikjl},
\quad
j\in\delta_1^+(i)
\qquad
\ba{l}
i=1,2,\dots,n1-1,\\
k=1,2,\ldots,n2-1
\ea
\end{equation}
\begin{equation}\label{const2}
x_{ik}\ge\sum_{l \in \delta_2^-(k)} y_{jlik},
\quad
j\in\delta_1^-(i)
\qquad
\ba{l}
 i=2,3,\dots,n1,\\
 k=2,3,\ldots,n2
\ea
\end{equation}
\begin{equation}\label{const3}
x_{ik} \ge\sum_{j \in \delta_1^+(i)} y_{ikjl},
\quad
l\in\delta_2^+(k)
\qquad
\ba{l}
i=1,2,\dots,n1-1,\\
k=1,2,\ldots,n2-1
\ea
\end{equation}
\begin{equation}\label{const4}
x_{ik}\ge\sum_{j \in \delta_1^-(i)} y_{jlik},
\quad
l\in\delta_2^-(k)
\qquad
\ba{l}
i=2,3,\dots,n1,\\
k=2,3,\ldots,n2. \ea
\end{equation}
\begin{equation}\label{xinx}
\ba{l} x \in X \ea
\end{equation}

Actually, we know how to represent $X$ with linear
constraints. Recalling the definition of feasible
path, (\ref{xinx}) is equivalent to
\begin{equation}
\sum_{l=1}^{k}x_{il} + \sum_{j=1}^{i-1}x_{jk} \leq 1, \quad
i=1,2,\ldots, n1, \quad k=1,2,\ldots, n2.
\end{equation}

We recall  that from the definition of the feasible paths in $B'$
(non-crossing matching in $B$) the $j$-th residue from $P1$ could
be matched with at most one residue from $P2$ and vice-versa. This
explains the sums into right hand side of (\ref{const1}) and
(\ref{const3}) -- for arcs having their tails at vertex $(i,k)$;
and (\ref{const2}) and  (\ref{const4})-- for arcs heading to
$(i,k)$. Any $(i,k)(j,l)$ arc can be activated ($y_{ikjl}=1$)  iff 
$x_{ik}=1$ and $x_{jl}=1$ and in this case the respective
constraints are active because of the objective function.

A tighter description of the polytope defined by
(\ref{const1})--(\ref{const4}) and $0\le x_{ik}\le 1$, $0\le
y_{ikjl}$ could be obtained by lifting the constraints
(\ref{const2})   and (\ref{const4})  as it is shown in
Fig. \ref{fig:path} b). The points shown are just  the predecessors
of $(i,k)$ in graph $B'$ and they form a grid of $\delta_1^-(i)$
rows and $\delta_2^-(k)$ columns.  Let $i_1,i_2,\dots,i_s$ be all
the vertices in  $\delta_1^-(i)$ ordered according the numbering
of the vertices in $V_1$ and likewise $k_1,k_2,\dots,k_t$ in
$\delta_2^-(k)$. Then the vertices in the $l$-th column
$(i_1,k_l),(i_2,k_l),\dots (i_s,k_l)$ correspond to pairwise
crossing matching and at most one of them could be chosen in any
feasible solution $x\in X$ (see (\ref {const4})). This "all
crossing" property will stay even if we add to this set the
following two sets: $(i_1,k_1),(i_1,k_2),\dots,(i_1,k_{l-1})$ and
$(i_s,k_{l+1}),(i_s,k_{l+2}),\dots,(i_s,k_t)$. Denote by
$col_{ik}(l)$ the union of these three sets and analogously by
$row_{ik}(j)$ the corresponding union for the $j$-th row of the
grid. When the grid is one column/row  only the set
$row_{ik}(j)$/$col_{ik}(l)$  is empty.

Now a tighter LP relaxation of (\ref{const1})--(\ref{const4}) is obtained by
changing (\ref{const2}) with
\begin{equation}\label{const2a}
x_{ik}\ge\sum_{(r,s)\in row_{ik}(j)} y_{rsik},
\quad
j\in\delta_1^-(i)
\qquad
\ba{l}
i=2,3,\dots,n1,\\
k=2,3,\ldots,n2
\ea
\end{equation}
and (\ref{const4}) with
\begin{equation}\label{const4a}
x_{ik}\ge\sum_{(r,s) \in col_{ik}(l)} y_{rsik},
\quad
l\in  \delta_2^-(k)
\qquad
\ba{l}
i=2,3,\dots,n1,\\
k=2,3,\ldots,n2.
\ea
\end{equation}

Remark: Since we are going to apply the Lagrangian technique there
is no need neither for an explicit description of the set $X$
neither for lifting the constraints (\ref{const1}) (\ref{const3}).

\section{Lagrangian relaxation approach}\label{sec:lagri}

Here, we show how the Lagrangian relaxation of constraints (\ref{const2a}) and
(\ref{const4a}) leads to an efficiently solvable problem, yielding upper and
lower bounds that are generally better than those found by the best known exact
algorithm \cite{1001}.

Let $\lambda_{ikj}^h \geq 0$ (respectively $\lambda_{ikj}^v \geq
0$) be a Lagrangian multiplier assigned to each constraint
(\ref{const2a}) (respectively (\ref{const4a})). By adding  the
slacks of these constraints to the objective function with weights
$\lambda$, we obtain the Lagrangian relaxation of the CMO problem
\begin{equation}\label{obj-lagr}
\begin{array}{ll}
LR(\lambda)= & \displaystyle \max \sum_{i,k,j\in \delta_1^-(i)}\lambda_{ikj}^h(x_{ik} -\sum_{(r,s) \in row_{ik}(j)} y_{rsik})\\
& \displaystyle + \sum_{i,k,l\in  \delta_2^-(k)}\lambda_{ikl}^v(x_{ik}-\sum_{(r,s) \in col_{ik}(l)} y_{rsik})+
\sum_{(ik)(jl)\in E_{B'}} y_{ikjl}
\end{array}
\end{equation}
subject to $x\in X$, (\ref{const1}), (\ref{const3}) and $y\geq0$.

\begin{prop}
$LR(\lambda)$ can be solved in  $O(|V'| + |E'|  )$ time.
\end{prop}

Proof: For each $(i,k)\in V'$, if $x_{ik}=1$ then the optimal choice
$y_{ikjl}$ amounts to solving the following : The heads of all
arcs in $E'$ outgoing from $(i,k)$ form a
$|\delta^+(i)|\times|\delta^+(k)|$ table. To each point $(j,l)$ in
this table, we assign the profit $\max\{0,c_{ikjl}(\lambda)$\},  
where $c_{ikjl}(\lambda)$ is the coefficient of $y_{ikjl}$ in
(\ref{obj-lagr}). Each vertex in this table is a head of an arc
outgoing from $(i,k)$. Then the subproblem we need to solve
consists in finding a subset of these arcs having a maximal sum
$c_{ik}(\lambda)$ of profits(the arcs of negative weight are
excluded as a candidates for the optimal solution) and such that
their heads lay on a feasible path. This could be done by a
dynamic programming approach in $O(|\delta^+(i)||\delta^+(k)|)$
time. Once profits $c_{ik}(\lambda)$ have been computed for all
$(i,k)$ we can find the optimal solution to $LR(\lambda)$  by
using the same DP algorithm  but this time on the table of
$n1\times n2$ points with profits for $(i,k)$-th one given by
\begin{equation}
c_{ik}(\lambda) + \sum_{j\in \delta_1^-(i) } \lambda_{ikj}^h +
\sum_{l \in \delta_2^-(k)  } \lambda_{ikl}^v.
\end{equation}
where the last two terms are the coefficients of $x_{ik}$ in
(\ref{obj-lagr}).

 Remark: The inclusion $x\in X$ is explicitly incorporated in the DP
 algorithm.
 
\subsection{The algorithm}\label{sec:algo}

In order to find the tightest upper bound on $ v(CMO)$ (or
eventually to solve the problem), we need to solve
 in the dual
space of the Lagrangian multipliers $LD=\min_{\lambda\geq 0}
LR(\lambda)$, whereas $LR(\lambda)$ is a problem in $x,y$.  A number
of methods have been proposed to solve Lagrangian duals:
subgradient method, dual ascent methods,constraint generation
method, column generation, bundel methods,augmented Lagrangian
methods, etc. Here, we choose the subgradient method.  It is an
iterative method in which at iteration $t$, given the current
multiplier vector $\lambda^t$, a step is taken along a subgradient
of $LR(\lambda)$, then if necessary, the resulting point is
projected onto the nonnegative orthant.  It is well known that
practical convergence of the subgradient method is unpredictable.
For some problems, convergence is quick and fairly reliable, while
other problems tend to produce erratic behavior of the multiplier
sequence, or the Lagrangian value, or both.  In a "good" case, one
usually observe a saw-tooth pattern in the Lagrangian value for
the first iterations, followed by a roughly monotonic improvement
and asymptotic convergence to a value that is hopefully the
optimal Lagrangian bound. The computational runs on a  reach set
of real-life instances confirm a "good" case belonging of our
approach at some expense in the speed of the convergence.

In our realization,  the update scheme for
$\lambda_{ikj}$ is
$\lambda_{ikj}^{t+1}=\max \{0, \lambda_{ikj}^{t} - \Theta^t
g_{ikj}^{t}\}$ \footnote{analogously for $\lambda_{ikl}$}, where  $g_{ikj}^{t}=\bar{x}_{ik}- \sum \bar{y}_{jlik}$ (see (\ref{const2a})  and   (\ref{const4a})  for the sum definition)  is  the sub-gradient component ($0,1$,or $-1$), calculated on the
optimal solution $ \bar{x}$, $\bar{y}$ of $LR(\lambda^t)$. The
step size $\Theta^t$ is
$\Theta^t=\frac{\alpha(LR(\lambda^t)-Z_{lb})}{\sum
(g_{ikj}^t)^2+\sum(g_{ikl}^t)^2} $ where $Z_{lb}$ is a known lower
bound for the CMO problem and $\alpha$ is an input parameter.
 Into this approach the $x$-components  of $LR(\lambda^t)$ solution
  provides a feasible solution to CMO
and thus a lower bound also. The best one (incumbent) so far
obtained is used for fathoming the nodes whose upper bound falls
below the incumbent and also  in section \ref{sec:results} for
reporting the final gap. If $LD \leq v(CMO)$  then the problem is
solved. If $LD>v(CMO)$ holds, in order to obtain the optimal
solution, one could pass to a branch\&bound algorithm suitably
tailored for such an upper bounds generator.

\begin{figure}[ht]
\centering
\epsfig{figure=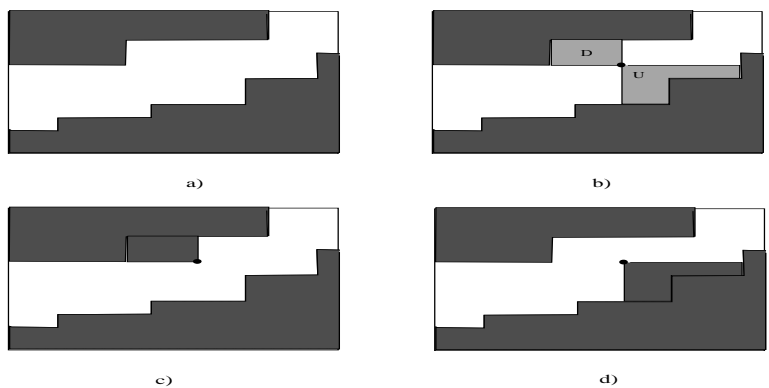, width=10cm, height=5cm}
\caption{Sketch of the B\&B   splitting strategy.  
~a)  the white  area in-between broken lines   represents the current node feasible set; 
~b) This set is split by $(rowbest,colbest)$, D corresponds to the set $S_d(rowbest,colbest)$
while U corresponds to the set $S_u(rowbest,colbest)$;  ~c) and d) are the descendants  of 
the node a). }\label{fig:bandb}
\end{figure}

From among various possible nodes splitting rules, the one shown
in Fig.~\ref{fig:bandb} gives quite satisfactory results (see
section~\ref{sec:results}). Formally,  let the current node be a subproblem of CMO defined over
the vertices of $V'$ falling in the interval $[lc(k),uc(k)]$ for
$k=1,n_2$ (in Fig.~\ref{fig:bandb} these are the points in-between
two broken lines (the white area). Let $(rowbest,colbest)$ be the $\arg\max
\min(S_u(i,k),S_d(i,k))$, where $S_d(i,k)=\sum_{j\leq
k}\max(uc(j)-i,0)$ and $S_u(i,k)=\sum_{j\geq k}\max(i-lc(j),0)$.
Now, the two descendants of the current node are obtained by
discarding from its feasible set the vertices in
$S_d(rowbest,colbest)$ and $S_u(rowbest,colbest)$ respectively.  The goal of
this strategy is twofold: to create descendants that are balanced
in sense of feasible set size and to reduce maximally the parent
node's feasible set.

 In addition, the following heuristics happened to be very
effective during the  traverse of  the B\&B tree nodes. Once the lower and the
upper bound are found at the root node,  an attempt to improve the lower bound
is realized as follows.

Let $(i_{k_1},k_1),(i_{k_2},k_2),\dots,(i_{k_s},k_s)$ be an
arbitrary feasible path which activates certain number of arcs
(recall that each iteration in the sub-gradient optimization phase
generates such path and lower bound as well).

 Then for a given strip size $sz$ (an input parameter set by
default to $4$), the matchings in the original CMO are restricted
to fall in a neighborhood of this path, allowing $x_{ik}$  to be
non zero only for
\[
\max\{1, i_j-sz \}\leq i \leq \min\{n1, i_j+sz \},
j=k_1,k_2,\dots,k_s.
\]

The Lagrangian dual of this subproblem is solved and a better
lower bound is possibly  sought. If the bound improves the
incumbent, the same procedure is repeated by changing the strip
alongside the new feasible solution.

Finally, the main steps of the B\&B algorithm are as follows:\\
\underline{Initialization:} Set $L$=\{original CMO problem, i.e. no restrictions on the feasible paths\}.\\
\underline{Problem selection and relaxation:} Select and delete the problem
$P^i$ from $L$ having the biggest upper bound. Solve the Lagrangian dual of
$P^i$. (Here  a repetitive call to a heuristics
is included after each improvement on the lower bound).\\
\underline{Fathoming and Pruning:} Follow classical rules.\\
\underline{Partitioning :} Create two  descendants of $P^i$ using $(rowbest,colbest)$ and add them to $L$.\\
\underline{Termination :} if $L=\emptyset$, the solution $(x^*,y^*)$ yielding the objective value is optimal.

\section{Numerical results}\label{sec:results}
To evaluate the above algorithm we performed two kinds of experiments. In the first one we  compared our approach with the best existing algorithm from literature  \cite{1001} in term of performance and quality of the bounds.
This comparison was done on a set of proteins suggested by Jeffrey Skolnick which was used in various recent papers related to protein structure comparison
\cite{1001,krasnogor,cmos_07}. This set contains 40 medium size domains from 33 proteins, which number of residues varies from 95 (\text{2b3iA}) to 252  (\text{1aw2A}). The maximum number of contacts is 593 (\text{1btmA}).
We afterwards experimentally evaluated the capability of our algorithm to perform as classifier on the Proteus\_300 set, a significantly larger protein set. It contains 300 domains, which number of residues varies from 64 (\text{d15bba\_}) to 455 (\text{d1po5a\_}). Its maximum number of contact is 1761 (\text{d1i24a\_}). We will soon make available all data and results\footnote{solved instances, upper and lower bounds, computational time, classifications...} on the URL:\\
{\em http://www.irisa.fr/symbiose/softwares/resources/proteus300}

\subsection{Performance and quality of bounds}
The results presented in this section were obtained on machines with AMD Opteron(TM) CPU at 2.4~GHz, 4 Gb Ram, RedHat 9 Linux. The algorithm was implemented in C.
According to SCOP classification\footnote{Using SCOP version 1.71} \cite{murzin}, the Skolnick set contains five families (see Table ~\ref{skolnick-set} in Annexe)\footnote{Caprara et al.  \cite{1001} mention only four families. 
This wrong classification was also accepted in \cite{krasnogor} but not in \cite{cmos_07}.
The families are in fact five as shown in Table ~\ref{skolnick-set}. According to SCOP classification the protein \text{1rn1} does not belong to the first family as
indicated in  \cite{1001}. Note that this corroborates the results obtained in \cite{1001} but the authors considered it as a mistake.}.
Note that both approaches that we compare use different Lagrangian relaxations.
Our algorithm is called  \verb?a_purva?\footnote{Apurva (Sanskrit) = not having existed before, unknown, wonderful, ...},  while the other Lagrangian algorithm is denoted by \verb?LR?.

The Skolnick set requires aligning 780 pairs of domains. We bounded the execution time to 1800 seconds for both algorithms.   \verb?a_purva? succeeded to solve 171 couples in the given time, while \verb?LR? solved  only 157 couples.
Note that another exact algorithm called \verb?CMOS? has been proposed in a very recent paper \cite{cmos_07}.  \verb?CMOS? succeeded to solve only 161 instances from the Skolnick set, 
yet the time limit was 4 hours on a similar workstation. Hence  it seems that 171 is the best score ever obtained when exactly solving Skolnick set.
To the best of our knowledge, we are the first ones to solve all the 164 instances with couples from the same SCOP folds, as well as the first to solve instances  with couples from different folds (the 7 instances of the $6^{th}$ class presented in Table \ref{table-1}). The interested reader can find our detailed results on the webpage cited before.

\begin{figure}[!ht]
\centering
\epsfig{figure=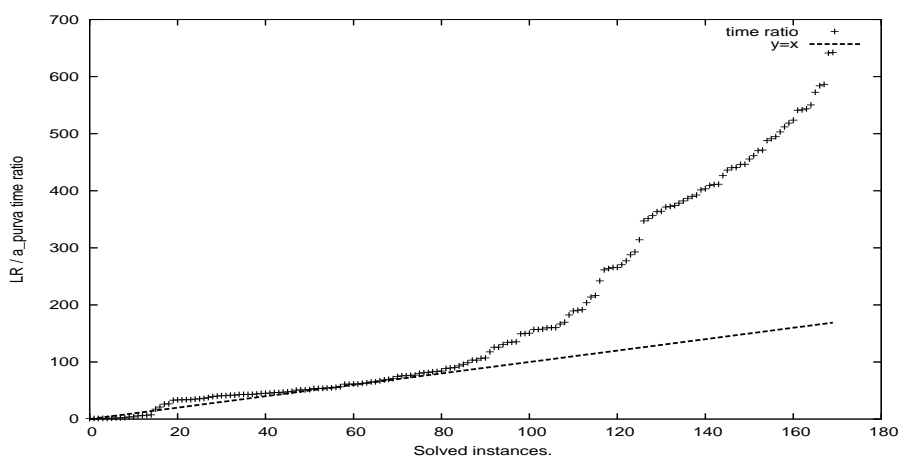, width=6cm, height=12cm,angle=-90}
\caption{$\frac{\mbox{LR time}}{\mbox{a\_purva time}}$ ratio as a function of solved instances}\label{fig:times}
\end{figure}

 Figure~\ref{fig:times} illustrates \verb?LR?/\verb?a_purva? time ratio  as a function of solved instances. It is easily seen that \verb?a_purva? is significantly faster than \verb?LR?  (up to several hundred times in the majority of cases). 
Table \ref{table-1}  in the Annexe contains more details concerning a subset of 164 pairs of proteins.  We observed that this set is a very interesting one.
It is characterized by the following properties: a) in all but the 6 last instances 
the \verb?a_purva? running time is less than 10 seconds; b) in all instances the relative gap\footnote{We define the relative gap as $100\times\frac{UB-LB}{UB}$.} at the root of the B\&B is smaller than 4, while in all other instances this gap is much larger (greater than 18 even for couples solved in less
than 1800 sec); c) this set contains all instances such that both proteins belong to the same family according to SCOP classification. In other words,  each pair such that both proteins belong to the same family is an easily solvable
instance for  \verb?a_purva? and this feature can be successfully used as a discriminator.  In fact, by virtue of this relation 
we were able to correctly classify the  40 items in the Skolnick set in 2000 seconds overall running time for all 780 instances. We will go back over this point in the next section.

\begin{figure}[!ht]
\begin{center} \epsfig{figure=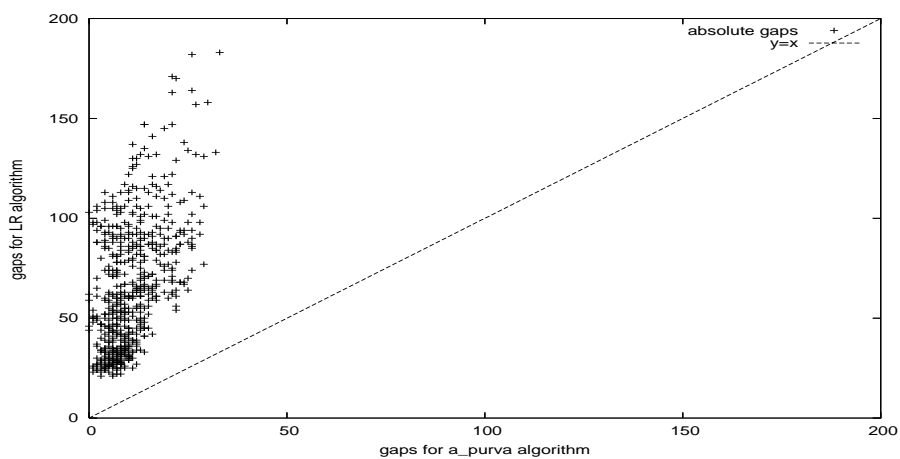, width=6cm, height=12cm,angle=-90}\end{center}
\caption{Comparing absolute gaps on the set of unsolved instances. The gaps computed by 
\text{a\_purva} are significantly smaller.}
\label{fig:gaps}
\end{figure}

Our next observation (see Fig.  ~\ref{fig:gaps} and Fig.~\ref{fig:bounds} in the Annexe) concerns the quality of gaps obtained by both algorithms on the set of unsolved instances. Remember that when a Lagrangian algorithm stops because of time limit (1800 sec. in our case) it provides two bounds: one upper (UB),  and one lower (LB).
Providing these bounds is a real advantage of a B\&B type algorithm compared to any meta-heuristics. These values can be used as a measure for how far is the optimization process from finding the exact optimum. The value \text{UB-LB} is usually called absolute gap. Any one of the 609 points $(x,y)$ in Fig. ~\ref{fig:gaps} presents the  absolute gap for \verb?a_purva? ($x$ coordinate) and for \verb?LR? ($y$ coordinate) algorithm.  All points are above the $y=x$  line (i.e. the absolute gap for  \verb?a_purva? is always smaller than the absolute gap for  \verb?LR?). On the other hand the entire figure is very asymmetric in a profit of our algorithm since its maximal  absolute gap is 33, while it is 183 for \verb?LR?.

In  Fig.~\ref{fig:bounds} we  similarly compare lower and upper bounds separately.   Any point  $\circ$ has the lower bound computed by \verb?a_purva? (res. \verb?LR?)  as $x$  (res. $y$) coordinate, while any point  $\times$ has the upper bound computed by \verb?a_purva? (res. \verb?LR?)  as $x$ (res. $y$) coordinate. We observe that in a large majority  the points   $\circ$ are below the $y=x$ line while the points $\times$ are above this line. This means that usually   \verb?a_purva? lowers bounds are higher, while its upper bounds are all smaller and therefore \verb?a_purva? provides bounds with clearly better quality than \verb?LR?. We don't have much information about the bounds find by \verb?CMOS?, except that at the root of the B\&B tree, it obtains upper bounds of worst quality than the ones of \verb?LR?.

\subsection{A\_purva as a classifier}
When running \verb?a_purva? on the Skolnick set, we observed that  relative gaps are smaller for similar domains than for dissimilar ones. This became even more obvious when we fixed a small upper bound of iterations and limited the computations only to the root of the B\&B tree.
The question then was to check if the relative gap can be used as a similarity   index
(the smaller is the relative gap, the more similar are the domains) which can be given to an automatic classifier in order to quickly provide a classification. 

We used the following protocol :
the runs of \verb?a_purva? were limited to the root, with a limit of 500 iterations for the subgradient descent. We used the publicly available hierarchical ascendant 
classifier \verb?Chavl? \cite{lerman}, 
which proposes a best partition of classified elements based on the derivative of the similarity index and thus requires no similarity threshold.
For the Skolnick set, the alignment of all couples was done in less than 1100 seconds (with a mean  computation time of 1.39 seconds/couple). The classification returned by \verb?Chavl? based on the relative gap is exactly the classification at the fold level in SCOP. Taking into account that according to  Table \ref{table-1}, 609 couples ran 1800 seconds without finding the solution,  this result  pushes to use the relative gap as a classifier.  Note also 
that we succeeded to classify the Skolnick set significantly faster than both previously  published exact algorithms \cite{cmos_07,1001} that use similarity indexes  based on lower bound only. This illustrates the effectiveness of using a similarity based on both upper and lower bounds.

To get a stronger confirmation of \verb?a_purva? classifier capabilities, we performed the same operation on the Proteus\_300 set, presented in Table \ref{table-2}.
The alignment\footnote{Detailed results of the runs will be available in our web page.} of the 44850 couples required roughly 82 hours (with a mean computation time of 6,58 seconds/couple).

Table \ref{table-3} presents the  classification that we obtain. It contains  25 classes denoted by letters A-Y. This classification is almost identical to the SCOP one (at folds level) which contains 24 classes denoted by numbers (presented in Table \ref{table-2}).
18 of the 24 SCOP classes correspond perfectly to our classes.
Class 15 (resp. 24) contains two families\footnote{In the SCOP classification, Families are sub-sub-classes of Folds.}  that we classified in M and  N (resp. V and  W).
Classes 9 and 11 were merged into class I and are indeed similar, with some domains (like \text{d1jgca\_} and \text{d1b0b\_\_}) having more than 75\% of common contacts\footnote{The percentage of common contacts between domains $i$ and $j$ is $\frac{CMO(i,j)}{MIN(C_{i},C_{j})}$ where $C_{i}$ (resp $C_{j}$) denotes the number of contacts in domain $i$ (resp $j$), and $CMO(i,j)$ is the number of common contacts between $i$ and $j$ found by a\_purva.}.
 Class 18 was split into its two families (X and Y), but Y was merged with class 10.  Again, some of the corresponding domains (e.g. \text{d1b00a\_} and \text{d1wb1a4}) are very  similar, with more than 75\% of common contacts.

\section{Conclusion}

In this paper, we give an efficient exact B\&B algorithm for contact map overlap problem. The bounds are found by using  Lagrangian relaxation and the dual problem is solved by sub-gradient approach. The efficiency of the algorithm is demonstrated on a benchmark set of 40 domains 
and the dominance over the existing algorithms is total. In addition,its capacity as classifier  (and this was the primary goal) was tested on a large data set of 300 protein domains. We were able to obtain in a short time a classification in  very good agreement to the well known SCOP database.

We are curently working on the integration of biological information into the contact maps, such as the secondary structure type of the residues (alpha helix or beta strand). Aligning only residues from the same type will reduce the research space and thus speed up the algorithm.

\section{Acknowledgement}

Supported by ANR grant Calcul Intensif projet  PROTEUS (ANR-06-CIS6-008)
and by Hubert Curien French-Bulgarian partnership ``RILA 2006'' N$^0$ 15071XF.

N. Malod-Dognin is supported by Région Bretagne.

We are thankful to Professor Giuseppe Lancia for numerous discussions and for kindly providing  
us with the source code and the contact map graphs for the Skolnick set.

All computations were done on the Ouest-genopole bioinformatics platform (http://genouest.org).

\input Biblio.tex
\appendix

\newpage
\begin{center}
{\bf ANNEXE}
  \end{center}
\input table

\input{Skolnick-set}

\begin{figure}[!ht]
\centerline{
    \epsfig{figure=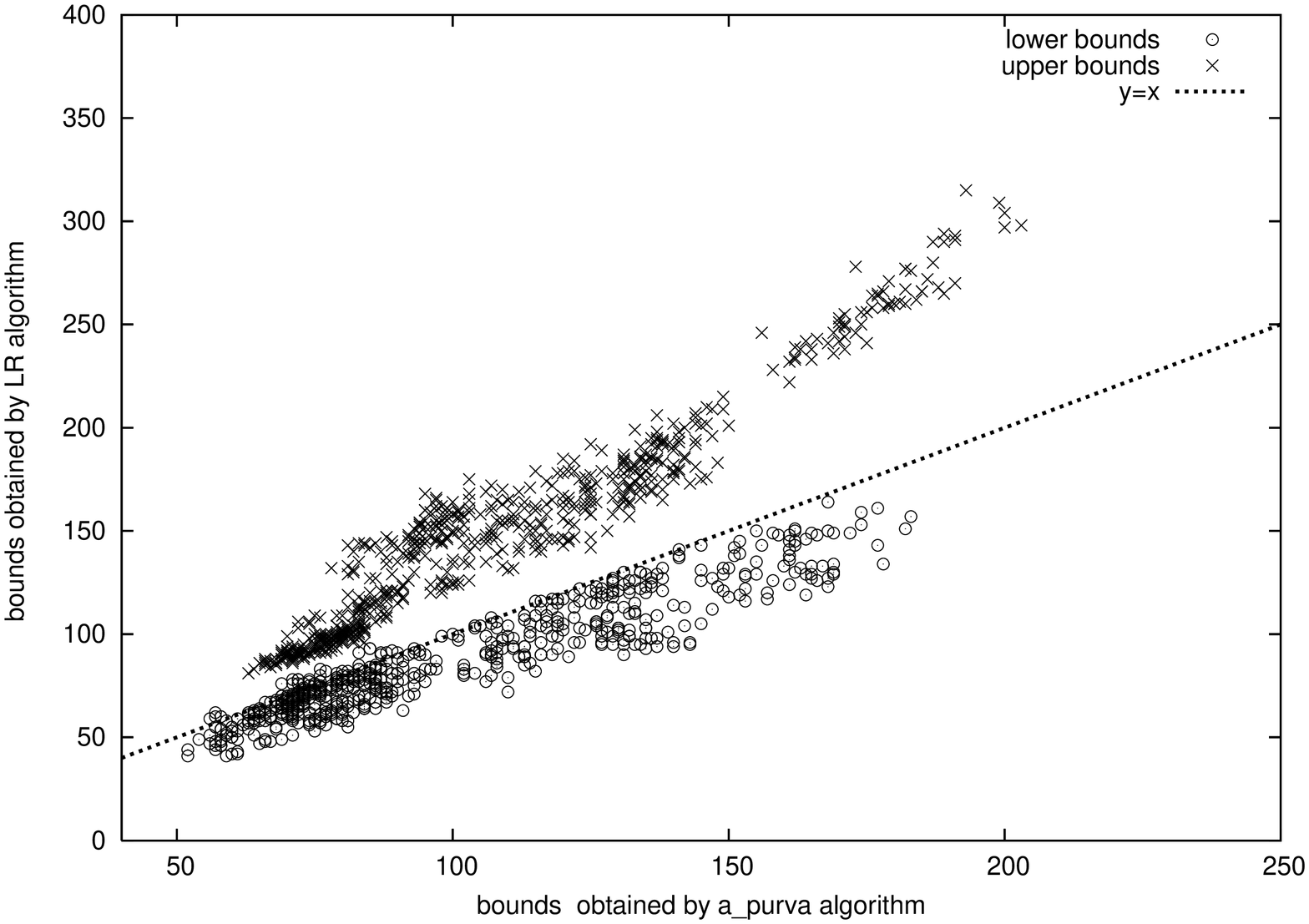, width=8cm, height=15cm, angle=-90}
}
\caption{Comparing the quality of lower and upper bounds on the set of unsolved instances. \text{a\_purva} clearly outperforms  \text{LR} on the quality of its bounds.}
\label{fig:bounds}
\end{figure}

\input t300_classif_SCOP.tex

\input t300_classif_gap.tex

\newpage

\end{document}

%% file: table
{\footnotesize

\begin{longtable}[htbp]{|l|l|r|r|r||l|r|r|r|}
\hline
F &Proteins      &CMO& Time       &Time   & Proteins  & CMO  & Time  &Time\\
  & Name         &    & LR         &a\_pr & Name      &      & LR    &a\_pr\\
\hline
1&1b00A~1dbwA&149&192.00&1.2&1ntr\_~1qmpA&119&545.94&7.18\\
1&1b00A~1nat\_&145&166.98&1.11&1ntr\_~1qmpB&115&454.01&4.23\\
1&1b00A~1ntr\_&118&565.47&3.59&1ntr\_~1qmpC&116&610.93&6.56\\
1&1b00A~1qmpA&143&198.72&1.33&1ntr\_~1qmpD&118&522.53&4.44\\
1&1b00A~1qmpB&136&439.95&59.65&1ntr\_~3chy\_&130&339.86&5.53\\
1&1b00A~1qmpC&139&263.81&1.68&1ntr\_~4tmyA&126&450.05&3.34\\
1&1b00A~1qmpD&137&181.23&1.89&1ntr\_~4tmyB&127&399.26&3.75\\
1&1b00A~3chy\_&154&141.50&0.85&1qmpA~1qmpB&221&3.77&0.03\\
1&1b00A~4tmyA&155&143.92&0.9&1qmpA~1qmpC&232&0.35&0.02\\
1&1b00A~4tmyB&155&75.41&0.73&1qmpA~1qmpD&230&0.02&0.03\\
1&1dbwA~1nat\_&157&226.42&1.51&1qmpA~3chy\_&160&69.78&1.07\\
1&1dbwA~1ntr\_&130&426.13&5.53&1qmpA~4tmyA&162&98.21&0.78\\
1&1dbwA~1qmpA&152&159.74&2.93&1qmpA~4tmyB&164&50.48&0.62\\
1&1dbwA~1qmpB&150&63.63&1.52&1qmpB~1qmpC&221&1.60&0.02\\
1&1dbwA~1qmpC&150&180.52&2.38&1qmpB~1qmpD&220&1.61&0.03\\
1&1dbwA~1qmpD&152&111.28&1.78&1qmpB~3chy\_&156&68.17&0.84\\
1&1dbwA~3chy\_&164&84.22&1.19&1qmpB~4tmyA&157&51.32&0.58\\
1&1dbwA~4tmyA&161&73.71&1.1&1qmpB~4tmyB&156&66.11&0.64\\
1&1dbwA~4tmyB&163&47.87&1.11&1qmpC~1qmpD&226&3.65&0.02\\
1&1nat\_~1ntr\_&127&302.39&3.59&1qmpC~3chy\_&157&75.14&1.23\\
1&1nat\_~1qmpA&157&66.03&1.04&1qmpC~4tmyA&162&55.46&1.26\\
1&1nat\_~1qmpB&149&69.00&0.99&1qmpC~4tmyB&162&78.52&0.58\\
1&1nat\_~1qmpC&152&73.53&1.07&1qmpD~3chy\_&158&59.47&1.11\\
1&1nat\_~1qmpD&151&99.14&1.33&1qmpD~4tmyA&157&59.23&0.71\\
1&1nat\_~3chy\_&163&76.95&0.86&1qmpD~4tmyB&159&53.27&0.59\\
1&1nat\_~4tmyA&175&15.58&0.28&3chy\_~4tmyA&171&54.33&0.55\\
1&1nat\_~4tmyB&172&19.06&0.37&3chy\_~4tmyB&174&41.43&0.5\\
1 &             &   &     &     &4tmyA~4tmyB&230&0.02&0.02\\
\hline
\hline
2&1bawA~1byoA&152&11.59&0.25&1byoB~2b3iA&135&7.21&0.27\\
2&1bawA~1byoB&155&6.11&0.18&1byoB~2pcy\_&175&2.28&0.05\\
2&1bawA~1kdi\_&140&33.84&0.55&1byoB~2plt\_&174&3.90&0.06\\
2&1bawA~1nin\_&153&9.45&0.21&1kdi\_~1nin\_&129&52.53&1.13\\
2&1bawA~1pla\_&124&28.04&0.62&1kdi\_~1pla\_&126&33.59&0.89\\
2&1bawA~2b3iA&130&15.57&0.38&1kdi\_~2b3iA&122&40.83&0.84\\
2&1bawA~2pcy\_&148&6.91&0.16&1kdi\_~2pcy\_&145&15.19&0.3\\
2&1bawA~2plt\_&161&5.22&0.13&1kdi\_~2plt\_&150&24.56&0.32\\
2&1byoA~1byoB&192&2.61&0.02&1nin\_~1pla\_&130&22.76&0.69\\
2&1byoA~1kdi\_&148&17.89&0.35&1nin\_~2b3iA&129&25.55&0.5\\
2&1byoA~1nin\_&140&30.14&0.85&1nin\_~2pcy\_&139&23.31&0.49\\
2&1byoA~1pla\_&150&7.55&0.16&1nin\_~2plt\_&146&18.85&0.52\\
2&1byoA~2b3iA&132&10.26&0.39&1pla\_~2b3iA&122&12.65&0.32\\
2&1byoA~2pcy\_&176&2.18&0.04&1pla\_~2pcy\_&143&4.75&0.14\\
2&1byoA~2plt\_&172&3.77&0.07&1pla\_~2plt\_&144&7.10&0.17\\
2&1byoB~1kdi\_&152&11.89&0.21&2b3iA~2pcy\_&127&11.79&0.35\\
2&1byoB~1nin\_&141&21.05&0.6&2b3iA~2plt\_&140&7.37&0.17\\
2&1byoB~1pla\_&148&6.94&0.16&2pcy\_~2plt\_&172&3.67&0.06\\
\hline
\hline
3&1amk\_~1aw2A&411&1272.28&1.48&1btmA~1tmhA&432&1801.97&2.81\\
3&1amk\_~1b9bA&400&1044.23&2.04&1btmA~1treA&433&1512.26&2.59\\
3&1amk\_~1btmA&427&1287.48&2.38&1btmA~1tri\_&419&1455.08&3.26\\
3&1amk\_~1htiA&407&265.16&1.4&1btmA~1ydvA&385&692.72&1.52\\
3&1amk\_~1tmhA&424&638.26&1.29&1btmA~3ypiA&406&1425.09&2.43\\
3&1amk\_~1treA&411&716.51&1.52&1btmA~8timA&408&940.59&2\\
3&1amk\_~1tri\_&445&447.54&0.97&1htiA~1tmhA&416&588.98&1.07\\
3&1amk\_~1ydvA&384&462.44&1.05&1htiA~1treA&426&395.23&0.81\\
3&1amk\_~3ypiA&412&427.66&0.97&1htiA~1tri\_&412&779.84&1.55\\
3&1amk\_~8timA&410&386.73&0.94&1htiA~1ydvA&382&405.04&1.09\\
3&1aw2A~1b9bA&411&961.04&3.28&1htiA~3ypiA&422&148.75&0.56\\
3&1aw2A~1btmA&434&750.67&3.1&1htiA~8timA&463&112.65&0.52\\
3&1aw2A~1htiA&425&363.03&1.78&1tmhA~1treA&513&119.27&0.23\\
3&1aw2A~1tmhA&474&185.72&0.51&1tmhA~1tri\_&413&630.57&2.19\\
3&1aw2A~1treA&492&157.79&0.37&1tmhA~1ydvA&384&785.56&1.5\\
3&1aw2A~1tri\_&408&1313.53&3.51&1tmhA~3ypiA&417&766.79&2.11\\
3&1aw2A~1ydvA&386&650.55&1.62&1tmhA~8timA&421&516.44&1.47\\
3&1aw2A~3ypiA&401&895.17&2.28&1treA~1tri\_&401&1169.41&2.68\\
3&1aw2A~8timA&423&276.06&1.76&1treA~1ydvA&389&1419.90&2.21\\
3&1b9bA~1btmA&441&653.29&2.08&1treA~3ypiA&407&522.65&1.34\\
3&1b9bA~1htiA&394&809.23&2.27&1treA~8timA&425&310.95&1.15\\
3&1b9bA~1tmhA&418&548.56&1.34&1tri\_~1ydvA&371&1040.31&1.92\\
3&1b9bA~1treA&410&613.99&1.25&1tri\_~3ypiA&412&607.52&1.75\\
3&1b9bA~1tri\_&391&1804.98&3.32&1tri\_~8timA&412&830.38&1.45\\
3&1b9bA~1ydvA&362&1608.97&6.1&1ydvA~3ypiA&374&355.82&0.92\\
3&1b9bA~3ypiA&396&700.45&1.88&1ydvA~8timA&388&399.47&0.99\\
3&1b9bA~8timA&392&634.48&1.66&3ypiA~8timA&418&267.14&0.65\\
3&1btmA~1htiA&403&1566.88&3.51&&&&\\
\hline
\hline
4&1b71A~1bcfA&211&1800.08&453.08&1bcfA~1rcd\_&222&528.84&1.99\\
4&1b71A~1dpsA&174&1800.43&266.54&1dpsA~1fha\_&180&1800.24&9.45\\
4&1b71A~1fha\_&216&1802.46&303.02&1dpsA~1ier\_&184&1800.31&8.42\\
4&1b71A~1ier\_&214&1801.32&480.43&1dpsA~1rcd\_&184&1490.02&5.7\\
4&1b71A~1rcd\_&211&1802.48&319&1fha\_~1ier\_&299&69.34&0.25\\
4&1bcfA~1dpsA&187&510.17&3.81&1fha\_~1rcd\_&295&36.40&0.19\\
4&1bcfA~1fha\_&218&1017.59&2.69&1ier\_~1rcd\_&297&24.03&0.15\\
4&1bcfA~1ier\_&226&556.33&3.28&&&&\\
\hline
\hline
5&1rn1A~1rn1B&191&1.23&0.03&1rn1B~1rn1C&197&0.21&0.01\\
5&1rn1A~1rn1C&190&1.01&0.03&&&&\\
\hline
\hline
6&1qmpD~1tri\_&131&1801.09&1674.98&1byoB~1rn1C&66&1800.09&686.03\\
6&1kdi\_~1qmpD&73&1800.15&904.75&1dbwA~1treA&145&1802.01&1703.2\\
6&1tmhA~4tmyB&112&1802.80&1521.23&1dbwA~1tri\_&149&1800.73&1173.5\\
6&1dpsA~4tmyB&89&1800.39&913.24&&&&\\
\hline
\hline
\caption{Column one contains the number of the families according to table \ref{skolnick-set}. 
The sixth class contains the hardest solved Skolnick set intstances. Column two(six) contains the names of the couples, column three(seven) is the score, column four(height) gives the time in seconds taken by LR algoritm, and column five(nine)  presents the corresponding time taken by \text{a\_purva}. }\label{table-1}
\end{longtable}
}

%% file: Skolnick-set
\begin{table}[!ht]
\begin{center}
\begin{tabular}{|c|l|l|l|}
\hline
 & Fold & Family &  Proteins \\
\hline
1 & Flavodoxin-like & CheY-related & \text{1b00, 1dbw, 1nat, 1ntr,} \\ 
 &  & & \text{ 1qmp(A,B,C,D), 3chy, 4tmy(A,B)} \\ 
 \hline
2 & Cupredoxin-like & Plastocyanin/ & \text{1baw, 1byo(A,B), 1kdi, 1nin, 1pla}\\
 &                   &    azurin-like              & \text{2b3i, 2pcy, 2plt}\\
 \hline
3 & TIM beta/alpha- & Triosephosphate & \text{1amk, 1aw2, 1b9b, 1btm, 1hti}\\
 &  barrel &  isomerase (TIM)  &  \text{1tmh, 1tre, 1tri, 1ydv, 3ypi, 8tim}\\
 \hline
4 &  Ferritin-like & Ferritin & \text{1b71, 1bcf, 1dps, 1hfa, 1ier, 1rcd}\\
\hline
5 &  Microbial & Fungal & \text{1rn1(A,B,C)}\\
 & ribonucleases   &  ribonucleases   & \\
\hline
\end{tabular}
\caption{The Skolnick set}\label{skolnick-set}
\end{center}
\end{table}

%% file: t300_classif_SCOP.tex
\begin{longtable}[htpb]{|c|l|l|l|}
            \hline
            {\tiny Fold number}&{\tiny SCOP fold}&{\tiny SCOP family}& {\tiny Domains name}\\
            \hline
            {\tiny 1}   &{\tiny \text{7-bladed beta-propeller}}     &{\tiny \text{WD40-repeat}}             &{\tiny \text{d1nr0a1, d1nexb2, d1k8kc\_, d1p22a2, d1erja\_}}\\
                              &                                                 &                                             &{\tiny \text{d1tbga\_, d1pgua2, d1gxra\_, d1pgua1, d1nr0a2}}\\
            \hline
            {\tiny 2}   &{\tiny \text{Acyl-CoA N-acyltransferases}} &{\tiny \text{N-acetyl transferase,}}   &{\tiny \text{d1nsla\_, d1qsta\_, d1vhsa\_, d1s3za\_, d1n71a\_}}\\
                              &{\tiny \text{(NAT)}}                       &{\tiny \text{NAT}}                     &{\tiny \text{d1tiqa\_, d1q2ya\_, d1ghea\_, d1ufha\_, d1vkca\_}}\\
            \hline
            {\tiny 3}   &{\tiny \text{Beta-Grasp}}                  &{\tiny \text{Ubiquitin-related}}       &{\tiny \text{d1wh3a\_, d1mg8a\_, d1xd3b\_, d1wm3a\_, d1wiaa\_}}\\
                              &{\tiny \text{(ubiquitin-like)}}            &                                             &{\tiny \text{d1v5oa\_, d1v86a\_, d1v6ea\_, d1wjna\_, d1wjua\_}}\\
            \hline
            {\tiny 4}   &{\tiny \text{C-type lectin-like}}          &{\tiny \text{C-type lectin domain}}    &{\tiny \text{ d1tdqb\_, d1e87a\_, d1kg0c\_, d1qo3c\_, d1sl4a\_}}\\
                              &                                                 &                                             &{\tiny \text{d1h8ua\_, d1tn3\_\_, d1jzna\_, d2afpa\_, d1byfa\_}}\\
            \hline
            {\tiny 5}   &{\tiny \text{Cytochrome P450}}             &{\tiny \text{Cytochrome P450}}         &{\tiny \text{d1jipa\_, d1izoa\_, d1x8va\_, d1io7a\_, d1jpza\_}}\\
                              &                                                 &                                             &{\tiny \text{d1po5a\_, d1lfka\_, d1n40a\_, d1n97a\_, d1cpt\_\_}}\\
            \hline
            {\tiny 6}   &{\tiny \text{DNA clamp}}                   &{\tiny \text{DNA polymerase}}          &{\tiny \text{d1b77a1, d1plq\_1, d1ud9a1, d1dmla2, d1plq\_2}}\\
                              &                                                 &{\tiny \text{processivity factor}}     &{\tiny \text{d1iz5a2, d1t6la1, d1dmla1, d1iz5a1, d1u7ba1}}\\
            \hline
            {\tiny 7}   &{\tiny \text{Enolase N-terminal}}          &{\tiny \text{Enolase N-terminal}}      &{\tiny \text{d1ec7a2, d1sjda2, d1r0ma2, d1wuea2, d1jpdx2}}\\
                              &{\tiny \text{domain-like}}                 &{\tiny \text{ domain-like}}            &{\tiny \text{d1rvka2, d1muca2, d1jpma2, d2mnr\_2, d1yeya2}}\\
            \hline
            {\tiny 8}   &{\tiny \text{Ferredoxin-like}}             &{\tiny \text{HMA, heavy metal-associated}} &{\tiny \text{d1fe0a\_, d1fvqa\_, d1aw0\_\_, d1mwya\_, d1qupa2}}\\
                              &                                                 &{\tiny \text{domain}}                      &{\tiny \text{d1osda\_, d1cc8a\_, d1sb6a\_, d1kqka\_, d1cpza\_}}\\
            \cline{3-4}       &                                                 &{\tiny \text{Canonical RBD}}               &{\tiny \text{d1no8a\_, d1wg1a\_, d1oo0b\_, d1fxla1, d1h6kx\_}}\\
                              &                                                 &                                                 &{\tiny \text{d1wg4a\_, d1sjqa\_, d1wf0a\_, d1l3ka2, d1whya\_}}\\
            \hline
            {\tiny 9}   &{\tiny \text{Ferritin-like}}               &{\tiny \text{Ferritin}}                &{\tiny \text{d1lb3a\_, d1vela\_, d1o9ra\_, d1jgca\_, d1vlga\_}}\\
                              &                                                 &                                             &{\tiny \text{d1tjoa\_, d1nf4a\_, d1jiga\_, d1ji4a\_, d1umna\_}}\\
            \hline
            {\tiny 10}  &{\tiny \text{Flavodoxin-like}}             &{\tiny \text{CheY-related}}            &{\tiny \text{d1krwa\_, d1mb3a\_, d1qkka\_, d1b00a\_, d1a04a2}}\\
                              &                                                 &                                             &{\tiny \text{d1w25a1, d1w25a2, d1oxkb\_, d1u0sy\_, d1p6qa\_}}\\
            \hline
            {\tiny 11}  &{\tiny \text{Globin-like}}                 &{\tiny \text{Globins}}                 &{\tiny \text{d1b0b\_\_, d1it2a\_, d1x9fc\_, d1h97a\_, d1q1fa\_}}\\
                              &                                                 &                                             &{\tiny \text{d1cqxa1, d1wmub\_, d1irda\_, d3sdha\_, d1gcva\_}}\\
            \hline
            {\tiny 12}  &{\tiny \text{Glutathione S-transferase}}   &{\tiny \text{Glutathione S-transferase}}   &{\tiny \text{d1oyja1, d1eema1, d1n2aa1, d2gsq\_1, d1f2ea1}}\\
                              &{\tiny \text{(GST), C-terminal domain}}    &{\tiny \text{(GST), C-terminal domain}}    &{\tiny \text{d1nhya1, d1r5aa1, d1m0ua1, d1oe8a1, d1k3ya1}}\\
            \hline
            {\tiny 13}  &{\tiny \text{Immunoglobulin-like}}         &{\tiny \text{Fibronectin type III}}    &{\tiny \text{d1uc6a\_, d1bqua1, d1n26a2, d2hft\_2, d1axib2}}\\
                              &{\tiny \text{beta-sandwich}}               &                                             &{\tiny \text{d1lwra\_, d1fyhb2, d1cd9b1, d1lqsr2, d1f6fb2}}\\
            \cline{3-4}       &                                                 &{\tiny \text{C1 set domains (antibody}}&{\tiny \text{d1l6xa1, d2fbjh2, d1k5nb\_, d1mjuh2, d1fp5a1}}\\
                              &                                                 &{\tiny \text{constant domain-like)}}   &{\tiny \text{d1uvqa1, d1rzfl2, d1mjul2, d3frua1, d1k5na1}}\\
            \cline{3-4}       &                                                 &{\tiny \text{I set domains}}           &{\tiny \text{d1gl4b\_, d1zxq\_2, d1iray3, d1biha3, d1p53a2}}\\
                              &                                                 &                                             &{\tiny \text{d1ev2e2, d1p53a3, d1ucta1, d1gsma1, d1rhfa2}}\\
            \hline
            {\tiny 14}  &{\tiny \text{LDH C-terminal domain-like}}  &{\tiny \text{Lactate \& malate dehydrogenases}}  &{\tiny \text{d1ojua2, d1llda2, d7mdha2, d1t2da2, d1gv1a2}}\\
                              &                                                 &{\tiny \text{C-terminal domain}}                 &{\tiny \text{d2cmd\_2, d1hyea2, d1ez4a2, d1hyha2, d1b8pa2}}\\
            \hline
            {\tiny 15}  &{\tiny \text{NAD(P)-binding}}              &{\tiny \text{LDH N-terminal}}          &{\tiny \text{d1uxja1, d2cmd\_1, d1o6za1, d1obba1, d1ldna1}}\\
                              &{\tiny \text{Rossmann-fold domains}}       &{\tiny \text{domain-like}}             &{\tiny \text{d1t2da1, d1b8pa1, d1hyea1, d1hyha1, d1s6ya1}}\\
            \cline{3-4}       &                                                 &{\tiny \text{Tyrosine-dependent}}      &{\tiny \text{d1db3a\_, d1sb8a\_, d1ek6a\_, d1xgka\_, d1ja9a\_}}\\
                              &                                                 &{\tiny \text{oxidoreductases}}         &{\tiny \text{d1i24a\_, d1gy8a\_, d1iy8a\_, d1vl0a\_, d1w4za\_}}\\
            \hline
            {\tiny 16}  &{\tiny \text{Ntn hydrolase-like}}          &{\tiny \text{Proteasome subunits}}     &{\tiny \text{d1rypg\_, d1rypd\_, d1rypl\_, d1rypa\_, d1rypb\_}}\\
                              &                                                 &                                             &{\tiny \text{d1q5qa\_, d1rypk\_, d1ryph\_, d1ryp1\_, d1rypi\_}}\\
            \hline
            {\tiny 17}  &{\tiny \text{Nuclear receptor}}            &{\tiny \text{Nuclear receptor}}        & {\tiny \text{d1nq7a\_, d1pzla\_, d1r1kd\_, d1t7ra\_, d1n46a\_}}\\
                              &{\tiny \text{ligand-binding domain}}       &{\tiny \text{ligand-binding domain}}   & {\tiny \text{d1pk5a\_, d1xpca\_, d1pq9a\_, d1pdua\_, d1xvpb\_}}\\
            \hline
            {\tiny 18}  &{\tiny \text{P-loop containing nucleoside}}&{\tiny \text{Extended AAA}}            &{\tiny \text{d1w5sa2, d1d2na\_, d1lv7a\_, d1fnna2, d1sxje2}}\\
                              &{\tiny \text{triphosphate hydrolases}}     &{\tiny \text{ATPase domain}}           &{\tiny \text{d1l8qa2, d1njfa\_, d1sxja2, d1ny5a2, d1r7ra3}}\\
            \cline{3-4}       &                                                 &{\tiny \text{G proteins}}              &{\tiny \text{d1r8sa\_, d1wb1a4, d1mkya2, d1kk1a3, d1ctqa\_}}\\
                              &                                                 &                                             &{\tiny \text{d1wf3a1, d1r2qa\_, d1i2ma\_, d1svia\_, d3raba\_}}\\
            \hline
            {\tiny 19}  &{\tiny \text{PDZ domain-like}}             &{\tiny \text{PDZ domain}}              &{\tiny \text{d1ihja\_, d1g9oa\_, d1qava\_, d1r6ja\_, d1m5za\_}}\\
                              &                                                 &                                             &{\tiny \text{d1l6oa\_, d1ujva\_, d1iu2a\_, d1n7ea\_, d1gm1a\_}}\\
            \hline
            {\tiny 20}  &{\tiny \text{Periplasmic binding}}         &{\tiny \text{L-arabinose binding}}     &{\tiny \text{d1sxga\_, d2dri\_\_, d1jyea\_, d1guda\_, d1jdpa\_}}\\
                              &{\tiny \text{protein-like I}}              &{\tiny \text{protein-like}}            &{\tiny \text{d1jx6a\_, d1byka\_, d1qo0a\_, d8abp\_\_, d1tjya\_}}\\
            \hline
            {\tiny 21}  &{\tiny \text{Periplasmic binding}}         &{\tiny \text{Phosphate binding}}       &{\tiny \text{d1xvxa\_, d1lst\_\_, d1y4ta\_, d1amf\_\_, d1ursa\_}}\\
                              &{\tiny \text{protein-like II}}             &{\tiny \text{protein-like}}            &{\tiny \text{d1i6aa\_, d1pb7a\_, d1ii5a\_, d1sbp\_\_, d1atg\_\_}}\\
            \hline
            {\tiny 22}  &{\tiny \text{PLP-dependent transferases}}  &{\tiny \text{AAT-like}}                &{\tiny \text{d1bw0a\_, d1toia\_, d1w7la\_, d1o4sa\_, d1m6sa\_}}\\
                              &                                                 &                                             &{\tiny \text{d1uu1a\_, d1v2da\_, d1u08a\_, d1lc5a\_, d1gdea\_}}\\
            \hline
            {\tiny 23}  &{\tiny \text{Protein kinase-like}}         &{\tiny \text{Protein kinases}}         &{\tiny \text{d1tkia\_, d1s9ja\_, d1k2pa\_, d1vjya\_, d1phk\_\_}}\\
                              &{\tiny \text{(PK-like)}}                   &{\tiny \text{catalytic subunit}}       &{\tiny \text{d1xkka\_, d1rdqe\_, d1fvra\_, d1u46a\_, d1uu3a\_}}\\
            \hline
            {\tiny 24}  &{\tiny \text{TIM beta/alpha-barrel}}       &{\tiny \text{Beta-glycanases}}         &{\tiny \text{d1xyza\_, d1bqca\_, d1bhga3, d1nofa2, d1ecea\_}}\\
                              &                                                 &                                             &{\tiny \text{d1qnra\_, d1foba\_, d1h1na\_, d1uhva2, d7a3ha\_}}\\
            \cline{3-4}       &                                                 &{\tiny \text{Class I aldolase}}        &{\tiny \text{d1n7ka\_, d1w3ia\_, d1vlwa\_, d1gqna\_, d1ub3a\_}}\\
                              &                                                 &                                             &{\tiny \text{d1l6wa\_, d1o5ka\_, d1sfla\_, d1p1xa\_, d1ojxa\_}}\\
            \hline
        \caption{Scop classification of the Proteus\_300 set.}\label{table-2}

\end{longtable}

%% file: t300_classif_gap.tex
\begin{longtable}[htpb]{|c|l|l|l|}
            \hline
            {\tiny Class name}&{\tiny SCOP fold}&{\tiny SCOP family}& {\tiny Domains name}\\
            \hline
            {\tiny A}   &{\tiny \text{7-bladed beta-propeller}}     &{\tiny \text{WD40-repeat}}             &{\tiny \text{d1nr0a1, d1nexb2, d1k8kc\_, d1p22a2, d1erja\_}}\\
                              &                                                 &                                             &{\tiny \text{d1tbga\_, d1pgua2, d1gxra\_, d1pgua1, d1nr0a2}}\\
            \hline
            {\tiny B}   &{\tiny \text{Acyl-CoA N-acyltransferases}} &{\tiny \text{N-acetyl transferase,}}   &{\tiny \text{d1nsla\_, d1qsta\_, d1vhsa\_, d1s3za\_, d1n71a\_}}\\
                              &{\tiny \text{(NAT)}}                       &{\tiny \text{NAT}}                     &{\tiny \text{d1tiqa\_, d1q2ya\_, d1ghea\_, d1ufha\_, d1vkca\_}}\\
            \hline
            {\tiny C}   &{\tiny \text{Beta-Grasp}}                  &{\tiny \text{Ubiquitin-related}}       &{\tiny \text{d1wh3a\_, d1mg8a\_, d1xd3b\_, d1wm3a\_, d1wiaa\_}}\\
                              &{\tiny \text{(ubiquitin-like)}}            &                                             &{\tiny \text{d1v5oa\_, d1v86a\_, d1v6ea\_, d1wjna\_, d1wjua\_}}\\
            \hline
            {\tiny D}   &{\tiny \text{C-type lectin-like}}          &{\tiny \text{C-type lectin domain}}    &{\tiny \text{ d1tdqb\_, d1e87a\_, d1kg0c\_, d1qo3c\_, d1sl4a\_}}\\
                              &                                                 &                                             &{\tiny \text{d1h8ua\_, d1tn3\_\_, d1jzna\_, d2afpa\_, d1byfa\_}}\\
            \hline
            {\tiny E}   &{\tiny \text{Cytochrome P450}}             &{\tiny \text{Cytochrome P450}}         &{\tiny \text{d1jipa\_, d1izoa\_, d1x8va\_, d1io7a\_, d1jpza\_}}\\
                              &                                                 &                                             &{\tiny \text{d1po5a\_, d1lfka\_, d1n40a\_, d1n97a\_, d1cpt\_\_}}\\
            \hline
            {\tiny F}   &{\tiny \text{DNA clamp}}                   &{\tiny \text{DNA polymerase}}          &{\tiny \text{d1b77a1, d1plq\_1, d1ud9a1, d1dmla2, d1plq\_2}}\\
                              &                                                 &{\tiny \text{processivity factor}}     &{\tiny \text{d1iz5a2, d1t6la1, d1dmla1, d1iz5a1, d1u7ba1}}\\
            \hline
            {\tiny G}   &{\tiny \text{Enolase N-terminal}}          &{\tiny \text{Enolase N-terminal}}      &{\tiny \text{d1ec7a2, d1sjda2, d1r0ma2, d1wuea2, d1jpdx2}}\\
                              &{\tiny \text{domain-like}}                 &{\tiny \text{ domain-like}}            &{\tiny \text{d1rvka2, d1muca2, d1jpma2, d2mnr\_2, d1yeya2}}\\
            \hline
            {\tiny H}   &{\tiny \text{Ferredoxin-like}}             &{\tiny \text{HMA, heavy metal-associated}} &{\tiny \text{d1fe0a\_, d1fvqa\_, d1aw0\_\_, d1mwya\_, d1qupa2}}\\
                              &                                                 &{\tiny \text{domain}}                      &{\tiny \text{d1osda\_, d1cc8a\_, d1sb6a\_, d1kqka\_, d1cpza\_}}\\
            \cline{3-4}       &                                                 &{\tiny \text{Canonical RBD}}               &{\tiny \text{d1no8a\_, d1wg1a\_, d1oo0b\_, d1fxla1, d1h6kx\_}}\\
                              &                                                 &                                                 &{\tiny \text{d1wg4a\_, d1sjqa\_, d1wf0a\_, d1l3ka2, d1whya\_}}\\
            \hline
            {\tiny I}         &{\tiny \text{Ferritin-like}}                     &{\tiny \text{Ferritin}}                      &{\tiny \text{d1lb3a\_, d1vela\_, d1o9ra\_, d1jgca\_, d1vlga\_}}\\
                              &                                                 &                                             &{\tiny \text{d1tjoa\_, d1nf4a\_, d1jiga\_, d1ji4a\_, d1umna\_}}\\
            \cline{2-4}       &{\tiny \text{Globin-like}}                       &{\tiny \text{Globins}}                       &{\tiny \text{d1b0b\_\_, d1it2a\_, d1x9fc\_, d1h97a\_, d1q1fa\_}}\\
                              &                                                 &                                             &{\tiny \text{d1cqxa1, d1wmub\_, d1irda\_, d3sdha\_, d1gcva\_}}\\
            \hline
            {\tiny J}  &{\tiny \text{Glutathione S-transferase}}   &{\tiny \text{Glutathione S-transferase}}   &{\tiny \text{d1oyja1, d1eema1, d1n2aa1, d2gsq\_1, d1f2ea1}}\\
                              &{\tiny \text{(GST), C-terminal domain}}    &{\tiny \text{(GST), C-terminal domain}}    &{\tiny \text{d1nhya1, d1r5aa1, d1m0ua1, d1oe8a1, d1k3ya1}}\\
            \hline
            {\tiny K}  &{\tiny \text{Immunoglobulin-like}}         &{\tiny \text{Fibronectin type III}}    &{\tiny \text{d1uc6a\_, d1bqua1, d1n26a2, d2hft\_2, d1axib2}}\\
                              &{\tiny \text{beta-sandwich}}               &                                             &{\tiny \text{d1lwra\_, d1fyhb2, d1cd9b1, d1lqsr2, d1f6fb2}}\\
            \cline{3-4}       &                                                 &{\tiny \text{C1 set domains (antibody}}&{\tiny \text{d1l6xa1, d2fbjh2, d1k5nb\_, d1mjuh2, d1fp5a1}}\\
                              &                                                 &{\tiny \text{constant domain-like)}}   &{\tiny \text{d1uvqa1, d1rzfl2, d1mjul2, d3frua1, d1k5na1}}\\
            \cline{3-4}       &                                                 &{\tiny \text{I set domains}}           &{\tiny \text{d1gl4b\_, d1zxq\_2, d1iray3, d1biha3, d1p53a2}}\\
                              &                                                 &                                             &{\tiny \text{d1ev2e2, d1p53a3, d1ucta1, d1gsma1, d1rhfa2}}\\
            \hline
            {\tiny L}  &{\tiny \text{LDH C-terminal domain-like}}  &{\tiny \text{Lactate \& malate dehydrogenases}}  &{\tiny \text{d1ojua2, d1llda2, d7mdha2, d1t2da2, d1gv1a2}}\\
                              &                                                 &{\tiny \text{C-terminal domain}}                 &{\tiny \text{d2cmd\_2, d1hyea2, d1ez4a2, d1hyha2, d1b8pa2}}\\
            \hline
            {\tiny M}         &{\tiny \text{NAD(P)-binding}}              &{\tiny \text{LDH N-terminal}}          &{\tiny \text{d1uxja1, d2cmd\_1, d1o6za1, d1obba1, d1ldna1}}\\
                              &{\tiny \text{Rossmann-fold domains}}       &{\tiny \text{domain-like}}             &{\tiny \text{d1t2da1, d1b8pa1, d1hyea1, d1hyha1, d1s6ya1}}\\
            \hline
            {\tiny N}         &{\tiny \text{NAD(P)-binding}}              &{\tiny \text{Tyrosine-dependent}}      &{\tiny \text{d1db3a\_, d1sb8a\_, d1ek6a\_, d1xgka\_, d1ja9a\_}}\\
                              &{\tiny \text{Rossmann-fold domains}}       &{\tiny \text{oxidoreductases}}         &{\tiny \text{d1i24a\_, d1gy8a\_, d1iy8a\_, d1vl0a\_, d1w4za\_}}\\
            \hline
            {\tiny O}  &{\tiny \text{Ntn hydrolase-like}}          &{\tiny \text{Proteasome subunits}}     &{\tiny \text{d1rypg\_, d1rypd\_, d1rypl\_, d1rypa\_, d1rypb\_}}\\
                              &                                                 &                                             &{\tiny \text{d1q5qa\_, d1rypk\_, d1ryph\_, d1ryp1\_, d1rypi\_}}\\
            \hline
            {\tiny P}  &{\tiny \text{Nuclear receptor}}            &{\tiny \text{Nuclear receptor}}        & {\tiny \text{d1nq7a\_, d1pzla\_, d1r1kd\_, d1t7ra\_, d1n46a\_}}\\
                              &{\tiny \text{ligand-binding domain}}       &{\tiny \text{ligand-binding domain}}   & {\tiny \text{d1pk5a\_, d1xpca\_, d1pq9a\_, d1pdua\_, d1xvpb\_}}\\
            \hline
            {\tiny Q}  &{\tiny \text{PDZ domain-like}}             &{\tiny \text{PDZ domain}}              &{\tiny \text{d1ihja\_, d1g9oa\_, d1qava\_, d1r6ja\_, d1m5za\_}}\\
                              &                                                 &                                             &{\tiny \text{d1l6oa\_, d1ujva\_, d1iu2a\_, d1n7ea\_, d1gm1a\_}}\\
            \hline
            {\tiny R}  &{\tiny \text{Periplasmic binding}}         &{\tiny \text{L-arabinose binding}}     &{\tiny \text{d1sxga\_, d2dri\_\_, d1jyea\_, d1guda\_, d1jdpa\_}}\\
                              &{\tiny \text{protein-like I}}              &{\tiny \text{protein-like}}            &{\tiny \text{d1jx6a\_, d1byka\_, d1qo0a\_, d8abp\_\_, d1tjya\_}}\\
            \hline
            {\tiny S}  &{\tiny \text{Periplasmic binding}}         &{\tiny \text{Phosphate binding}}       &{\tiny \text{d1xvxa\_, d1lst\_\_, d1y4ta\_, d1amf\_\_, d1ursa\_}}\\
                              &{\tiny \text{protein-like II}}             &{\tiny \text{protein-like}}            &{\tiny \text{d1i6aa\_, d1pb7a\_, d1ii5a\_, d1sbp\_\_, d1atg\_\_}}\\
            \hline
            {\tiny T}  &{\tiny \text{PLP-dependent transferases}}  &{\tiny \text{AAT-like}}                &{\tiny \text{d1bw0a\_, d1toia\_, d1w7la\_, d1o4sa\_, d1m6sa\_}}\\
                              &                                                 &                                             &{\tiny \text{d1uu1a\_, d1v2da\_, d1u08a\_, d1lc5a\_, d1gdea\_}}\\
            \hline
            {\tiny U}  &{\tiny \text{Protein kinase-like}}         &{\tiny \text{Protein kinases}}         &{\tiny \text{d1tkia\_, d1s9ja\_, d1k2pa\_, d1vjya\_, d1phk\_\_}}\\
                              &{\tiny \text{(PK-like)}}                   &{\tiny \text{catalytic subunit}}       &{\tiny \text{d1xkka\_, d1rdqe\_, d1fvra\_, d1u46a\_, d1uu3a\_}}\\
            \hline
            {\tiny V}  &{\tiny \text{TIM beta/alpha-barrel}}       &{\tiny \text{Beta-glycanases}}         &{\tiny \text{d1xyza\_, d1bqca\_, d1bhga3, d1nofa2, d1ecea\_}}\\
                              &                                                 &                                             &{\tiny \text{d1qnra\_, d1foba\_, d1h1na\_, d1uhva2, d7a3ha\_}}\\
            \hline
            {\tiny W}  &{\tiny \text{TIM beta/alpha-barrel}}       &{\tiny \text{Class I aldolase}}        &{\tiny \text{d1n7ka\_, d1w3ia\_, d1vlwa\_, d1gqna\_, d1ub3a\_}}\\
                              &                                                 &                                             &{\tiny \text{d1l6wa\_, d1o5ka\_, d1sfla\_, d1p1xa\_, d1ojxa\_}}\\
            \hline

            {\tiny X}  &{\tiny \text{P-loop containing nucleoside}}&{\tiny \text{Extended AAA}}            &{\tiny \text{d1w5sa2, d1d2na\_, d1lv7a\_, d1fnna2, d1sxje2}}\\
                              &{\tiny \text{triphosphate hydrolases}}           &{\tiny \text{ATPase domain}}           &{\tiny \text{d1l8qa2, d1njfa\_, d1sxja2, d1ny5a2, d1r7ra3}}\\
            \hline
            {\tiny Y}         &{\tiny \text{P-loop containing nucleoside}}      &{\tiny \text{G proteins}}              &{\tiny \text{d1r8sa\_, d1wb1a4, d1mkya2, d1kk1a3, d1ctqa\_}}\\
                              &{\tiny \text{triphosphate hydrolases}}           &                                       &{\tiny \text{d1wf3a1, d1r2qa\_, d1i2ma\_, d1svia\_, d3raba\_}}\\
            \cline{2-4}       &{\tiny \text{Flavodoxin-like}}                   &{\tiny \text{CheY-related}}            &{\tiny \text{d1krwa\_, d1mb3a\_, d1qkka\_, d1b00a\_, d1a04a2}}\\
                              &                                                 &                                       &{\tiny \text{d1w25a1, d1w25a2, d1oxkb\_, d1u0sy\_, d1p6qa\_}}\\
            \hline

        \caption{Relative gap based classification of the Proteus\_300 set. Column 2 and 3 present the SCOP classification of the elements inside each classes}\label{table-3}

\end{longtable}

%% file: RR-6370.bbl
\begin{thebibliography}{10}

\bibitem{ABY04}
R.~Andonov, S.~Balev, and N.~Yanev.
\newblock Protein threading: From mathematical models to parallel implementations.
\newblock {\em INFORMS Journal on Computing}, 16(4), 2004.

\bibitem{Bal04}
S.~Balev.
\newblock Solving the protein threading problem by lagrangian relaxation.
\newblock In {\em Proceedings of WABI 2004: 4th Workshop on Algorithms in
  Bioinformatics}, LNCS/LNBI. Springer-Verlag, 2004.

\bibitem{cost}
P. ~Veber, N. ~Yanev, R. ~Andonov, V. ~Poirriez.
\newblock Optimal protein threading by cost-splitting.
\newblock {\em Lecture Notes in Bioninformatics}, 3692, pp.365-375, 2005

\bibitem{jcomp}
 N. ~Yanev, P. ~Veber, R. ~Andonov and S. ~Balev.
\newblock Lagrangian approaches for a class of matching problems.
 \newblock {\em INRIA PI 1814}, 2006 and in
\newblock {\em Journal of computational and applied mathematics}, 2007 (to appear)

\bibitem{1001}
A.~Caprara, R.~Carr, S.~Israil,  G.~Lancia and B.~Walenz.
\newblock 1001 Optimal PDB Structure Alignments: Integer Programming
Methods for Finding the Maximum Contact Map Overlap.
\newblock {\em Journal of  Computational Biology}, 11(1), 2004, pp. 27-52

\bibitem{101}
G.~Lancia, R.~Carr, and B.~Walenz.
\newblock 101 Optimal PDB Structure Alignments: A branch and cut algorithm for the Maximum Contact Map Overlap problem.
\newblock {\em RECOMB}, pp. 193-202, 2001

\bibitem{caprara}
A.~Caprara, and G.~Lancia.
\newblock Structural Alignment of Large-Size Protein via Lagrangian Relaxation.
\newblock {\em RECOMB}, pp. 100-108, 2002

\bibitem{lancia}
G.~Lancia, and S.~Istrail.
\newblock Protein Structure Comparison: Algorithms and Applications.
\newblock {\em Protein Structure Analysis and Desing}, pp. 1-33, 2003

\bibitem{strickland}
D.M.~Strickland, E.~Barnes, and J.S.~Sokol.
\newblock Optimal Protein Struture Alignment Using Maximum Cliques.
\newblock {\em OPERATIONS RESEARCH}, 53, 3, pp. 389-402, 2005

\bibitem{agarwal}
P.K.~Agarwal, N.H.~Mustafa, and Y.~Wang.
\newblock Fast Molecular Shape Matching Using Contact Maps.
\newblock {\em Journal of Computational Biology}, 14, 2, pp 131-147, 2007

\bibitem{krasnogor2}
N.~Krasnogor.
\newblock Self Generating Metaheuristic in Bioinformatics: The Proteins Structure Comparison Case.
\newblock {\em Genetic Programming and Evolvable Machines}, 5, pp 181-201, 2004

\bibitem{3djudge}
W.~Jaskowski, J.~Blazewicz, P.~Lukasiak, et al.
\newblock 3D-Judge - A Metaserver Approach to Protein Structure Prediction.
\newblock {\em Foundations of Computing and Decision Sciences}, 32, 1, 2007

\bibitem{cmo-hard}
D. ~Goldman, C.H. ~Papadimitriu, and S. ~Istrail.
\newblock Algorithmic aspects of protein structure similarity.
\newblock {\em FOCS 99: Proceedings of the 40th annual symposium on foundations of computer science} IEEE Computer Society,1999

\bibitem{xu-gang}
J. ~Xu, F. ~Jiao, B. ~Berger.
\newblock A parametrized Algorithm for Protein Structer Alignment.
\newblock {\em RECOMB 2006, Lecture Notes in Bioninformatics}, 3909,pp. 488-499,2006

\bibitem{halperin}
I. ~Halperin, B. ~Ma, H. ~Wolfson, et al.
\newblock Principles of docking: An overview of search algorithms and a guide to scoring functions.
\newblock {\em Proteins Struct. Funct. Genet.}, 47, 409-443,2002

\bibitem{goldman}
D. ~Goldman, S. ~Israil, C. ~Papadimitriu.
\newblock Algorithmic aspects of protein structure similarity.
\newblock {\em IEEE Symp. Found. Comput. Sci.} 512-522,1999

\bibitem{gary}
M. ~Garey, D. ~Johnson.
\newblock Computers and Intractability: A Guide to the Thoery of NP-completness.
\newblock {\em Freeman and company}, New york, 1979

\bibitem{krasnogor}
D. ~Pelta, N. ~Krasnogor, C. Bousono-Calzon, et al.
\newblock A fuzzy sets based generalization of contact maps for the overlap of protein structures.
\newblock {\em Journal of Fuzzy Sets and Systems, }  152(2):103-123, 2005.

\bibitem{murzin}
A.G. ~Murzin, S.E. ~Brenner, T. ~Hubbard and C. ~Chothiak.
\newblock SCOP: A structural classification of proteins database for the investigation of sequences and structures.
\newblock {\em Journal of Molecular Biology}, 247, pp. 536-540, 1995

\bibitem{lerman}
I.C. ~Lerman.
\newblock Likelihood linkage analysis (LLA) classification method (Around an example treated by hand).
\newblock {\em Biochimie, Elsevier éditions} 75, pp. 379-397, 1993

\bibitem{cmos_07}
W. ~Xie, and N. ~Sahinidis.
\newblock A Reduction-Based Exact Algorithm for the Contact Map Overlap Problem.
\newblock {\em Journal of Computational Biology}, V. 14, No 5, pp. 637-654, 2007


\bibitem{adam}
A.~Godzik.
\newblock The Structural alignment between two proteins: is there a unique answer?
\newblock {\em Protein Science}, 5, pp 1325-1338, 1996

\bibitem{astral}
J.-.M.~Chandonia, G.~Hon, N.S.~Walker, et al.
\newblock The ASTRAL Compendium in 2004.
\newblock {\em Nucleic Acids Reserch}, 32, 2004


\end{thebibliography}
